\documentclass[12pt]{iopart}

\usepackage{iopams,amssymb,epsfig,subfigure,graphicx,amsthm}

\usepackage{amscd}

\begin{document}
\title[Josephson tunneling of dark solitons in a double-well potential]{Josephson tunneling of dark solitons in a double-well potential}
\author{H.\ Susanto}
\address{School of Mathematical Sciences, University of Nottingham, University Park, Nottingham, NG7 2RD, UK}
\author{J.\ Cuevas}
\address{Grupo de F\'{\i}sica No Lineal. Universidad de Sevilla. Departamento de F\'{\i}sica Aplicada I. Escuela Polit\'enica Superior, C/ Virgen de Africa, 7, 41011 Sevilla, Spain}
\author{P.\ Kr\"uger}
\address{Midlands Ultracold Atom Research Centre, School of Physics and Astronomy, University of Nottingham, University Park, Nottingham, NG7 2RD, UK}

\begin{abstract}

We study the dynamics of matter waves in an effectively one-dimensional Bose-Einstein condensate in a double well potential. We consider in particular the case when one of the double wells confines excited states. Similarly to the known ground state oscillations, the states can tunnel between the wells experiencing the physics known for electrons in a Josephson junction, or be self-trapped. As the existence of dark solitons in a harmonic trap are continuations of such non-ground state excitations, one can view the Josephson-like oscillations as tunnelings of dark solitons. Numerical existence and stability analysis based on the full equation is performed, where it is shown that such tunneling can be stable. Through a numerical path following method, unstable tunneling is also obtained in different parameter regions. A coupled-mode system is derived and compared to the numerical observations. Regions of (in)stability of Josephson tunneling are discussed and highlighted.
Finally, we outline an experimental scheme designed to explore such dark soliton dynamics in the laboratory.

\end{abstract}



\maketitle

\newpage

\section{Introduction}

One fundamental physical phenomenon observable on a macroscopic scale is the Josephson tunneling (JT) of electrons between two superconductors connected by a weak link, predicted by Josephson in 1962 \cite{jose62}. It is due to the macroscopic wave functions with global phase coherence that have a small spatial overlap. The first observation of this effect was reported by Anderson et al.\ \cite{ande63}. 

Since the only requirement for the occurrence of JT is a weak coupling, other weakly connected macroscopic quantum samples were also expected to admit such tunneling. For neutral superfluids, JT has been observed in liquid $^3$He \cite{pere97} and $^4$He \cite{sukh01}. In the context of Bose-Einstein condensates (BECs) \cite{bose24,eins25,ande95,davi95,brad97,dalf99}, the prediction was made by Smerzi et al.\ \cite{smer97,ragh99,giov00}, followed by the experimental observation where a single \cite{albi05,levy07} and an array \cite{cata01} of short Bose-Josephson junctions (BJJs) were realized. The idea of BJJs has also been extended to a long BJJ \cite{kaur05,kaur06}, which mimics long superconducting Josephson junctions. Such a junction can be formed between two parallel quasi one-dimensional BECs linked by a weak coupling. Atomic Bose-Josephson vortices (BJVs) akin to Josephson fluxons in superconducting long Josephson junctions \cite{usti98} have been proposed as well \cite{kaur05,kaur06}. Moreover, it was emphasized that a BJV can transform from and to a dark soliton, due to the presence of a critical coupling at which the two solitonic structures exchange their stability.

The study of JT in BECs considers the tunneling of the Thomas-Fermi cloud, i.e.\ a continuation of the ground state. The tunneling dynamics has been explained using a two-mode approximation \cite{smer97,giov00}. The validity of the approximation has been shown in \cite{sacc03,sacc04}. To improve the applicability regime of such an approximation, modified coupled-mode equations have been presented in, e.g., \cite{ostr00,anan06,jia08,diaz10}.

It is important to note that in addition to the ground state, nonlinear excitations, such as dark matter waves, can also be created in BECs. Dark soliton dynamics  in BECs with single well potentials has been studied theoretically (see a review \cite{kono08}) and experimentally \cite{burg99,beck08,well08,stel08}. Interesting phenomena on the collective behavior of a quantum degenerate bosonic gas, such as soliton oscillations \cite{beck08,well08,theo10} and frequency shifts due to soliton collisions \cite{stel08} were observed. The evolution of solitons is of particular interest as the extent to which their behavior can be described in a particle picture is an open question and merits further experimental and theoretical investigation. A combination of soliton physics with the dynamics at weak links within double well potentials will shed light on the collective behavior of excited Bose-Einstein condensates in non-trivial potentials. In this paper, we present an analysis of the dynamics of dark matter waves in a double well potential. Static properties of such a configuration have been recently studied in \cite{midd10,ichi08}. Here, we show that dark matter waves can also experience stable quantum tunneling between the wells. This implies that localized excitations in higher dimensions, such as vortices, may also experience JT. The (in)stability is obtained using numerical Floquet analysis, which is applied for the first time in the study of JT. 
The numerical calculations are necessary as the stability of the observed tunneling is not immediately obvious. This is especially the case because dark solitons are higher-order excited states. The possibility that modes with lower energy will be excited is not ruled out by a coupled-mode approximation. 

The present paper is outlined as follows. In Section 2, we discuss the governing equation used in the current study. We then solve the equation numerically, where we obtain stable and unstable Josephson tunneling through a numerical path following method. The stability analysis is performed through calculating the Floquet multipliers of the solutions. In Section 3, we derive a coupled-mode approximation describing the tunneling dynamics. Good agreement between the numerics and the approximation is obtained and shown. 
In Section 4 we present a possible experimental setup to explore the results reported herein. Finally we conclude the work in Section 5.

\section{Josephson tunnelings}

\subsection{Mathematical model}

We consider the normalized nonlinear Schr\"odinger (NLS) equation modelling the BECs (see, e.g., \cite{carr08} for the scaling)
\begin{eqnarray}
i\psi_{t}+\psi_{xx}+s|\psi|^2\psi-V(x)\psi=0,
\label{ge1}
\end{eqnarray}
where $\psi$ is the bosonic field, and $t$ and $x$ is the time and position coordinate, respectively. The parameter $s=\pm1$ characterizes the attractive and repulsive nonlinear interaction, respectively, and $V(x)$ is the external double well potential, which for simplicity is taken as
\begin{equation}
V=\frac12\Omega^2(|x|-a)^2,
\label{ep}
\end{equation}
with the parameters $\Omega$ and $a$ controlling steepness and position of the two minima. The total number of atoms $N$ in the trap is conserved with
\begin{equation}N=\int_{-\infty}^{+\infty}|\psi|^2\,dx.\label{nor}\end{equation}
Throughout the present paper, we set $s=-1$, i.e.\ we consider repulsive interactions between particles.

For non-interacting particles ($s=0$) in a single well potential ($a=0$), the governing equation (\ref{ge1}) can be solved analytically to yield $\psi_n=e^{-iE_nt}\phi_n(x)$, where $\phi_n$ satisfies
\begin{equation}
\phi_{n+1}=(\frac{\Omega}{\sqrt[4]{2}} x-\frac{\sqrt[4]{2}}{\Omega}\partial_{x})\phi_n, \, n=0,1,2,\dots,
\label{linmod}
\end{equation}
with
\[
\phi_0=e^{-\frac{\Omega}{2\sqrt{2}}x^2},
\]
and the chemical potential $E_n$ is given by
\[E_n=\frac12\sqrt2(2n+1)\Omega.\]

The excitations $\phi_0$ can be continued to nonzero $s$, which has been considered in, e.g., \cite{vacc00,cole10,dodd96,kivs01,yuka97,tral08,kevr05,alfi07,zezy08,agos02_2,agos00} (see also \cite{agos02} for discussions on stationary solutions of the NLS equation with a multi-well potential that do not reduce to any of the eigenfunctions of the linear Schr\"odinger problem). The existence and the stability analysis of continuations of $\phi_n$ in a double-well potential has been presented in \cite{theo06}, where it was shown that there is a symmetry breaking of the corresponding solutions, i.e.\ a change of (in)stability from a symmetric to an asymmetric state. One typical manifestation of the instability is a periodic transfer of atoms between the wells, i.e.\ Josephson tunneling.

As most of Josephson tunneling studied in BECs considers the tunneling of the Thomas-Fermi cloud, which is a continuation of the ground state solution $\phi_0$, here we consider the tunneling of dark solitons, which can be viewed as continuations of excited states $\phi_{n>0}$.

\subsection{Numerical periodic solutions}

To look for solutions describing Josephson tunneling, we seek solutions that fulfills the relation $\psi(x,T)=\psi(x,0)$, with $T$ being the period of the Josephson oscillations. Such solutions posses double periodicity, i.e.\ one due to the solitonic nature with a period $2\pi/E$, where $E$ is the chemical potential (intra-well oscillations) and the other one caused by the Josephson effect (inter-well oscillations). Consequently, we can express the solutions in terms of a Fourier series multiplied by a factor related to the stationary character of dark solitons
\begin{equation}
    \psi(x,t)=\exp(-iEt)\sum_{k=-\infty}^{\infty} z_k(x)\exp(ik\omega t),
\end{equation}
where $\omega=2\pi/T$ is the Josephson oscillation frequency. These solutions are denoted as commensurate if the commensurability condition $E=(q/p)T=(2q\pi)/(\omega p)$ is fulfilled, with $\{q,p\}\in\mathbb{N}$. In what follows, we fix $p=1$.

Commensurate solutions are consequently fixed points of the map $\psi(x,0)\rightarrow\psi(x,T)$ and can be found either by using shooting methods in real space or algebraic methods in  Fourier space. In order to do that, we will transform the problem into a discrete one by means of a finite difference discretization with spatial step $\Delta x=0.2$ and apply the techniques developed for discrete breathers in Klein-Gordon lattices \cite{Marin,phason}. If a shooting method were used, a time step $\Delta t=0.02$ would be necessary. As the considered oscillations herein have periods about 1500 time units, this method would imply many integration steps. In addition to that, the lack of an analytical Jacobian would also imply the necessity of the numerical determination of this matrix. These facts suggest the suitability of the proposed Fourier space method, which, apart from transforming the set of differential equations into an algebraic one, provides an analytical expression for the Jacobian.

Truncating the Fourier series at $k_m$, i.e.\ the maximum value of $|k|$, which has been chosen to be 9 in most of the calculations due to computational reasons, Eq.\ (\ref{ge1}) yields a set of nonlinear equations with the $k$-th component of the dynamical equation set given by
\begin{equation}
    F_k(x)\equiv(E-\omega k)z_k+\partial^2_x z_k-V(x)z_k -s\sum_{m=-k_m}^{k_m}\sum_{n=-k_m}^{k_m}z_m z_n z_{k-m+n}=0.
\end{equation}
We then obtain the following expression for each component of the Jacobian
\begin{eqnarray}
    \frac{\partial F_k(x)}{\partial z_n(x')}&=&\left\{[E-\omega k-V(x)]\delta(x-x')+\partial^2_{xx}\right\}\delta_{k,n}\nonumber\\ &&-s\delta(x-x')\sum_m\left[z_m^*z_{k-n+m}+z_m(z_{k-m+n}+z^*_{n+m-k})\right],
\end{eqnarray}
where we have written $z_k\equiv z_k(x)$ in both equations.

\begin{figure}[tbph]
    \includegraphics[width=7cm,clip]{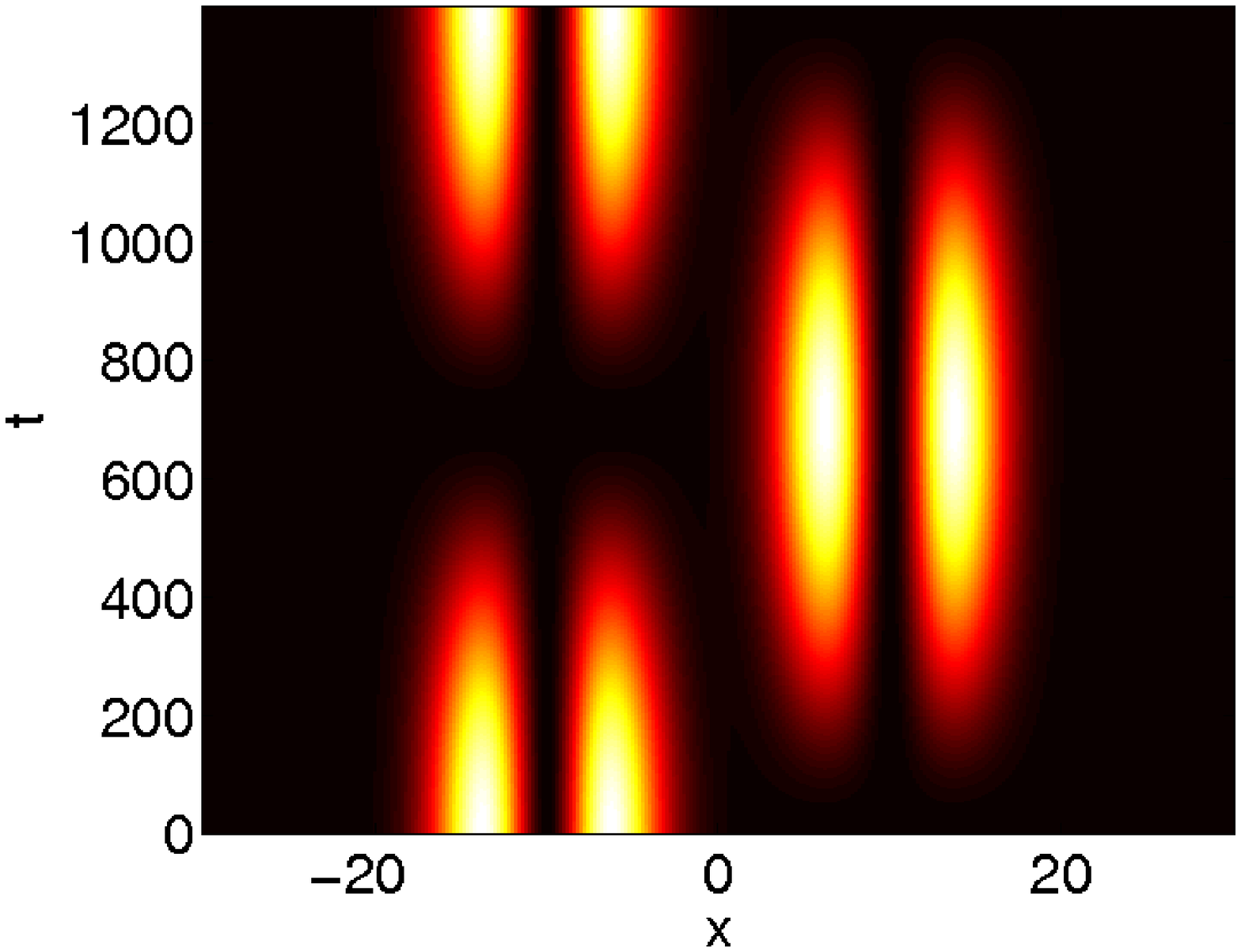}
    \includegraphics[width=7cm,clip]{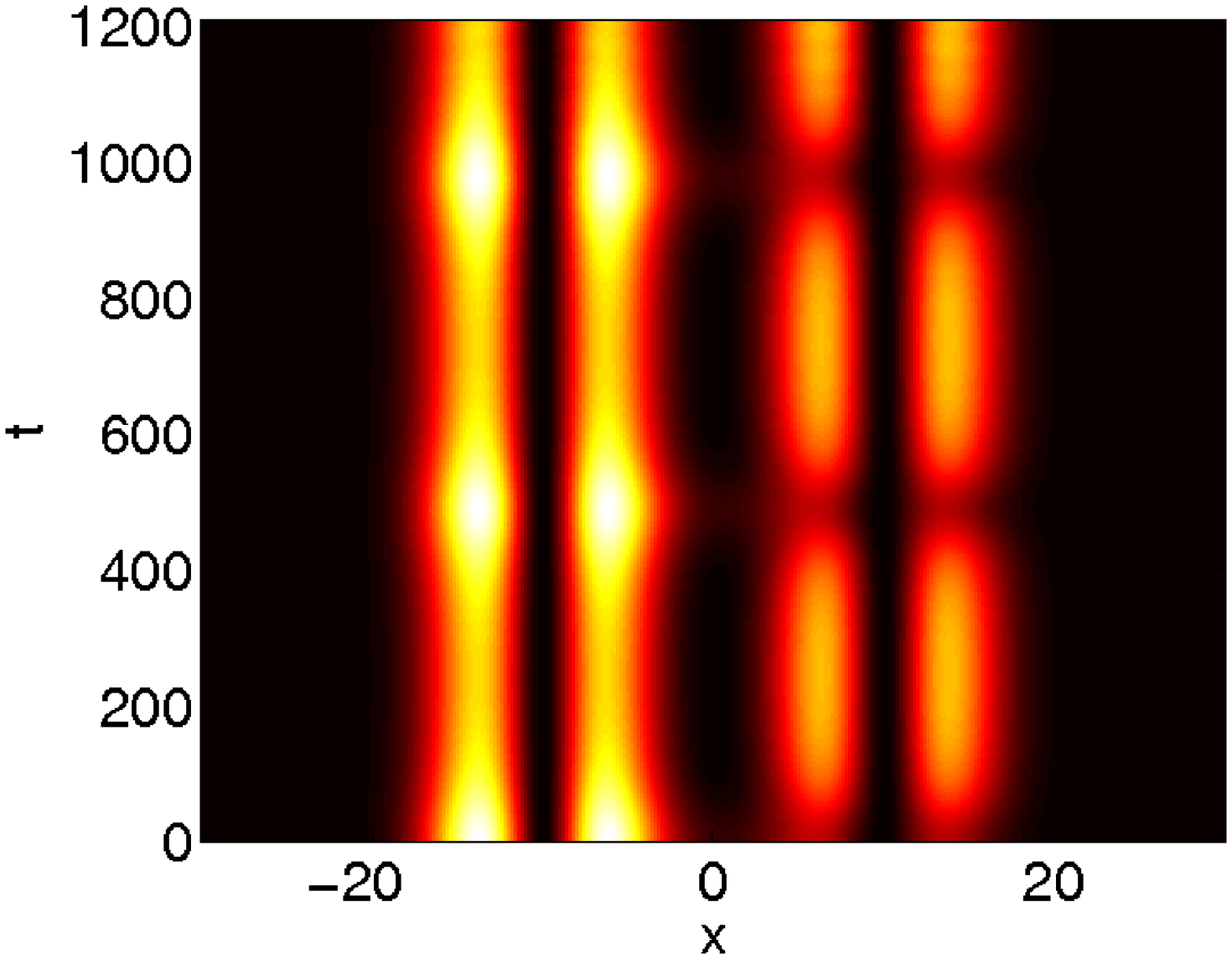}
         \includegraphics[width=7cm]{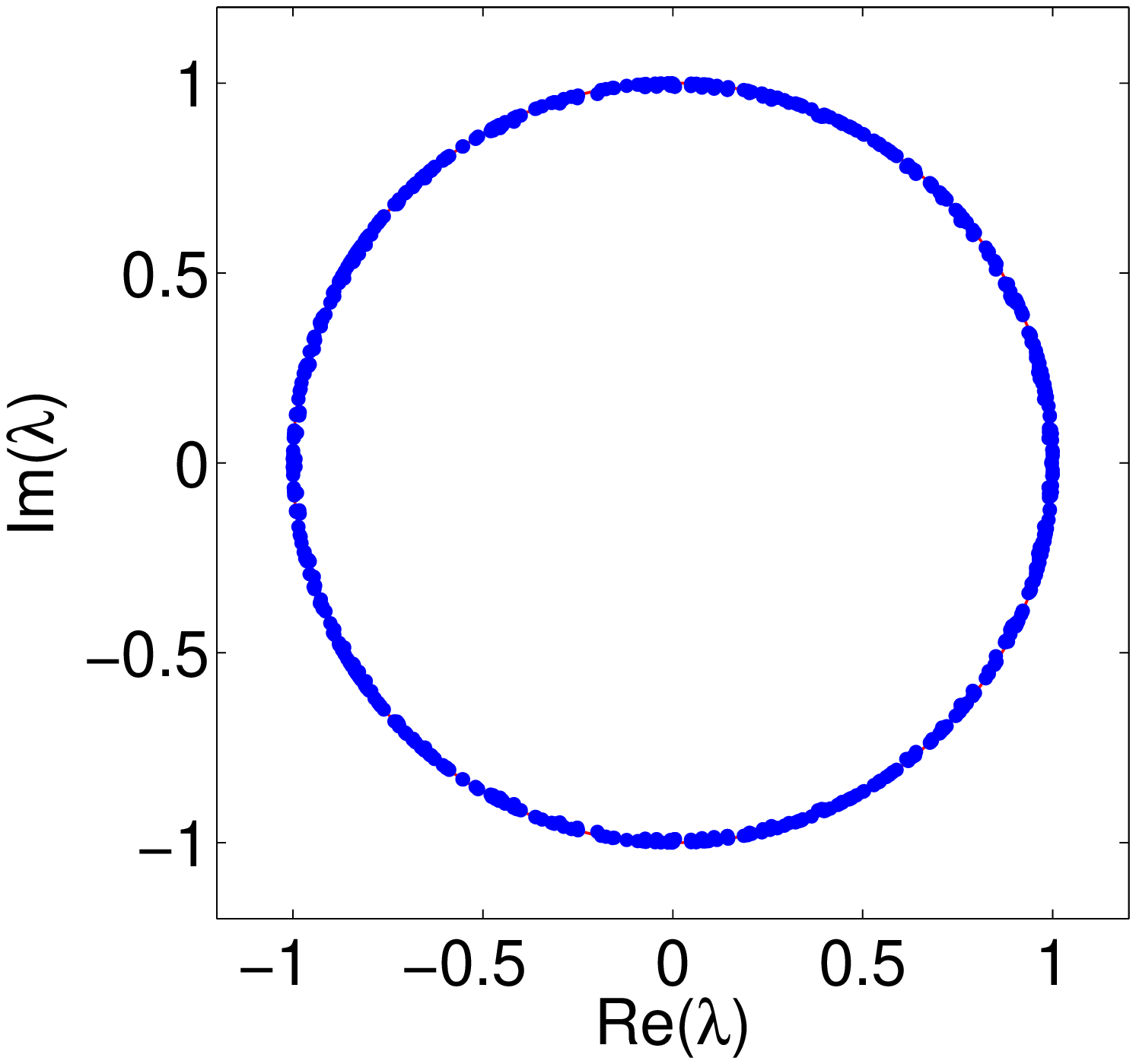}
    \includegraphics[width=7cm]{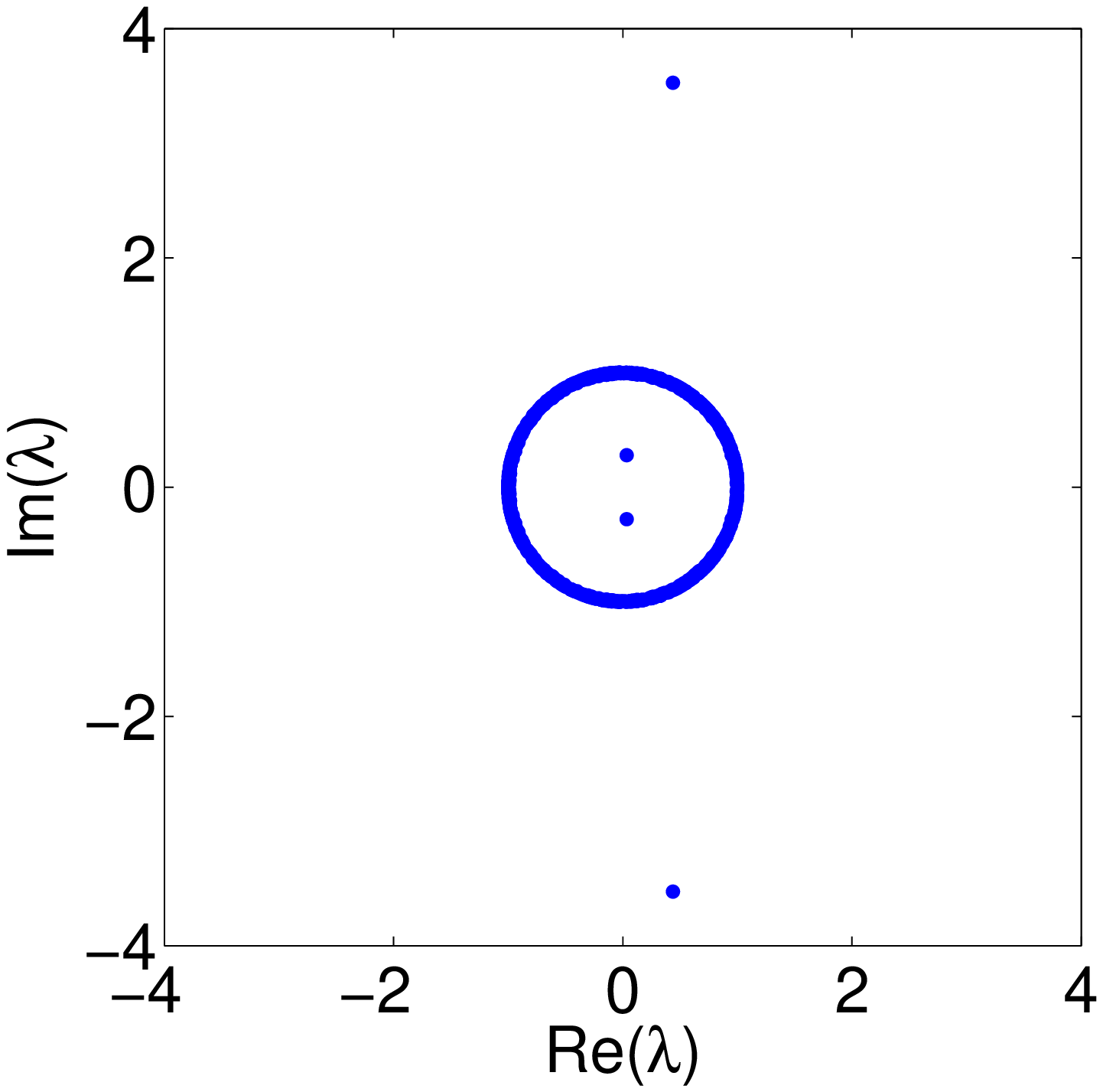}
    \includegraphics[width=7.5cm,clip]{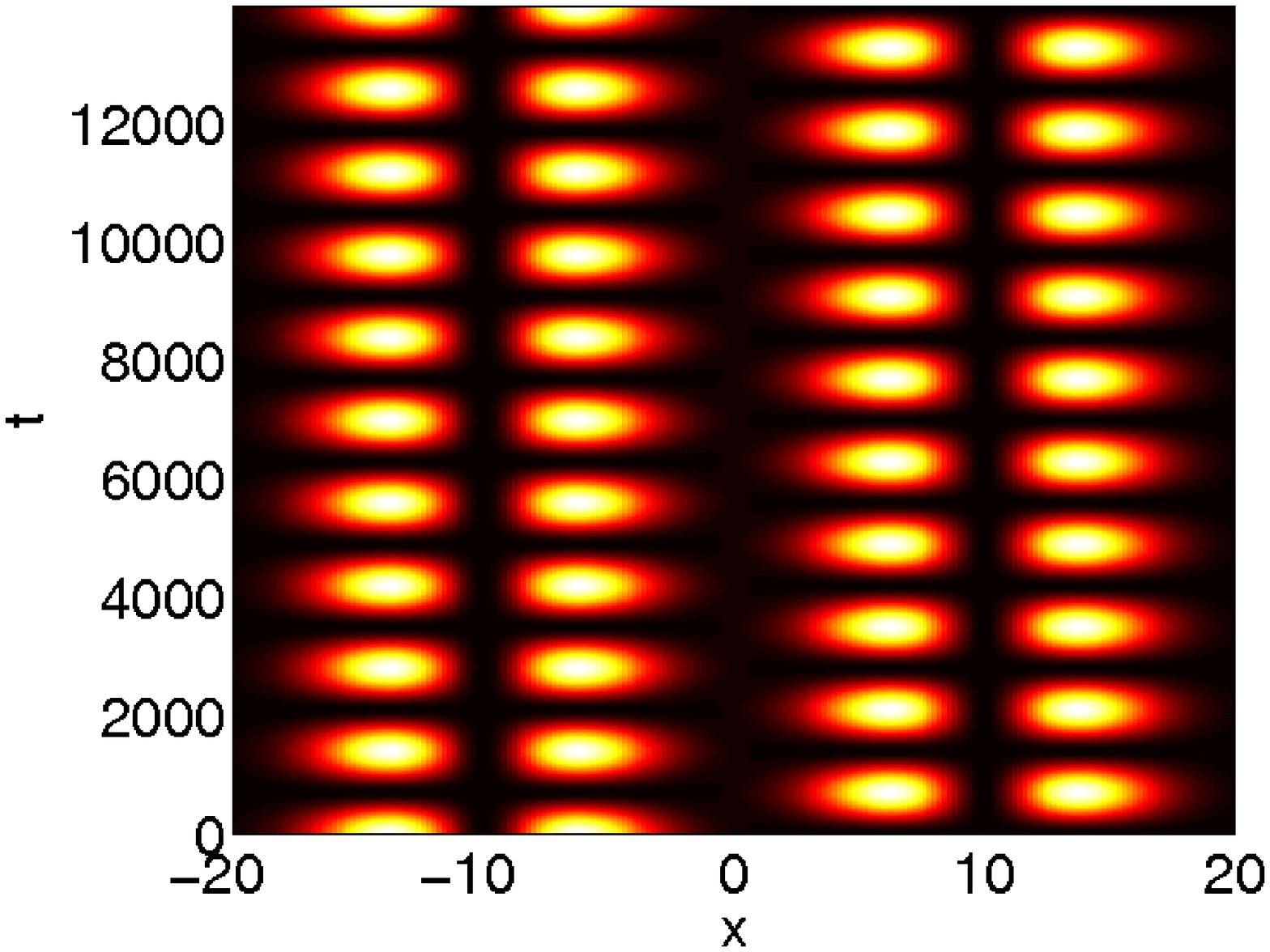}
    \includegraphics[width=7.5cm,clip]{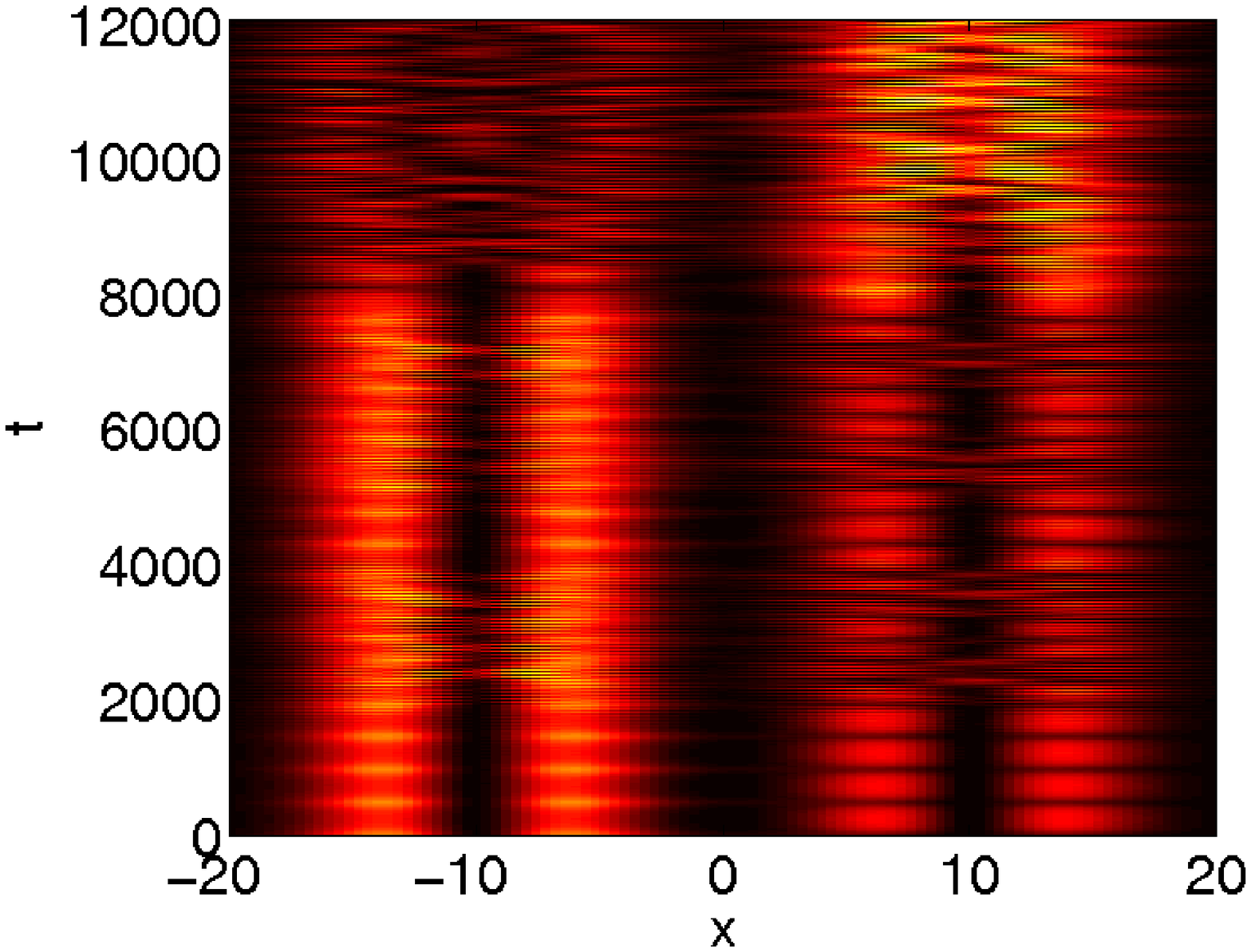}
\caption{(Top) The first few oscillations of the atom density $|\psi(x,t)|^2$ for dark solitons in a double well potential with $\Omega=0.1$, $a=10$, and (left) $\omega=0.00450$ 
and (right) $\omega=0.00520$, which respectively corresponds to $N=0.0340$ and $N=0.7677$. In both cases, the initial conditions are obtained from a numerical continuation with  $q=47$ (see the text). (Middle) Floquet multiplier distributions corresponding to solutions in the top left and right panel, respectively. (Bottom) Longer time evolutions of the top panels where one can see that the solution in the top right panel is indeed unstable. }
\label{fig:evol}
\end{figure}

Once a periodic solution, say $\Psi(x,t)$, is obtained, to study its (linear) orbital stability one needs to analyze the time evolution of a small perturbation $\xi(x,t)$ to $\Psi(x,t)$. The equation satisfied to leading order by $\xi(x,t)$ is
\begin{equation}\label{eq:stab}
    i\xi_t+\xi_{xx}-s(2|\Psi|^2\xi+\Psi^2\xi^*)-V(x)\xi=0.
\end{equation}
The stability properties are then determined by the spectrum of the Floquet operator $\mathcal{F}$ (whose matrix representation is the monodromy) defined as
\begin{equation}
    \left[\begin{array}{c} \mathrm{Re}(\xi(x,T)) \\ \mathrm{Im}(\xi(x,T)) \\ \end{array}
    \right]=\mathcal{F}\left[\begin{array}{c} \mathrm{Re}(\xi(x,0)) \\ \mathrm{Im}(\xi(x,0)) \\ \end{array}
    \right].
\end{equation}
A Floquet analysis can be performed as long as the solutions are commensurate.

As the Floquet operator is symplectic, it implies that there is always a pair of degenerate monodromy eigenvalues corresponding to the phase and growth modes at $1$. If the oscillations are stable, all the eigenvalues must lie on the unit circle (see \cite{Melvin} for a similar analysis in a discrete setting). In order to get the monodromy with enough accuracy, the simulations must be performed using a time step twenty times smaller than in the case of the dynamical equations, i.e.\ $\Delta t=0.001$ in this case.

We have calculated commensurate solitons for $\Omega=0.1$ and $a=10$ using the method described above and analysed the stability of those solutions. Presented in the top panels of Figure \ref{fig:evol} are two periodic solutions that we obtained together with the time evolution of a dark soliton in a double well potential. The left and right panel respectively corresponds to JT and a transition to macroscopic quantum self-trapping, similarly to the dynamics of the ground state oscillations \cite{smer97,giov00}.

In the middle panels of Figure \ref{fig:evol}, we present the distribution of the Floquet multipliers of the two solutions depicted in the top panels in the complex plane. Note that we did not obtain a continuum spectrum of the Floquet operator due to the discretization of the equations. It is worth noting that as there is a quartet of multipliers that do not lie on the unit circle, one can conclude that the solution in the top right panel is unstable. We show in the bottom panels of Figure \ref{fig:evol} a longer time evolution of the solutions in the top panels, where one can see that the solution in the top right panel is indeed unstable. The instability we reported here is a clear evidence that the nonlinearity term in the governing equation (\ref{ge1}) plays an important role, as all the solutions would have been stable otherwise. A typical instability dynamics is a repulsive interaction between the dark solitons in different wells so that they start to oscillate about the minimum of the wells. This can be clearly observed in the bottom right panel of Figure \ref{fig:evol}.

\begin{figure}[tbph]
    \includegraphics[width=7cm]{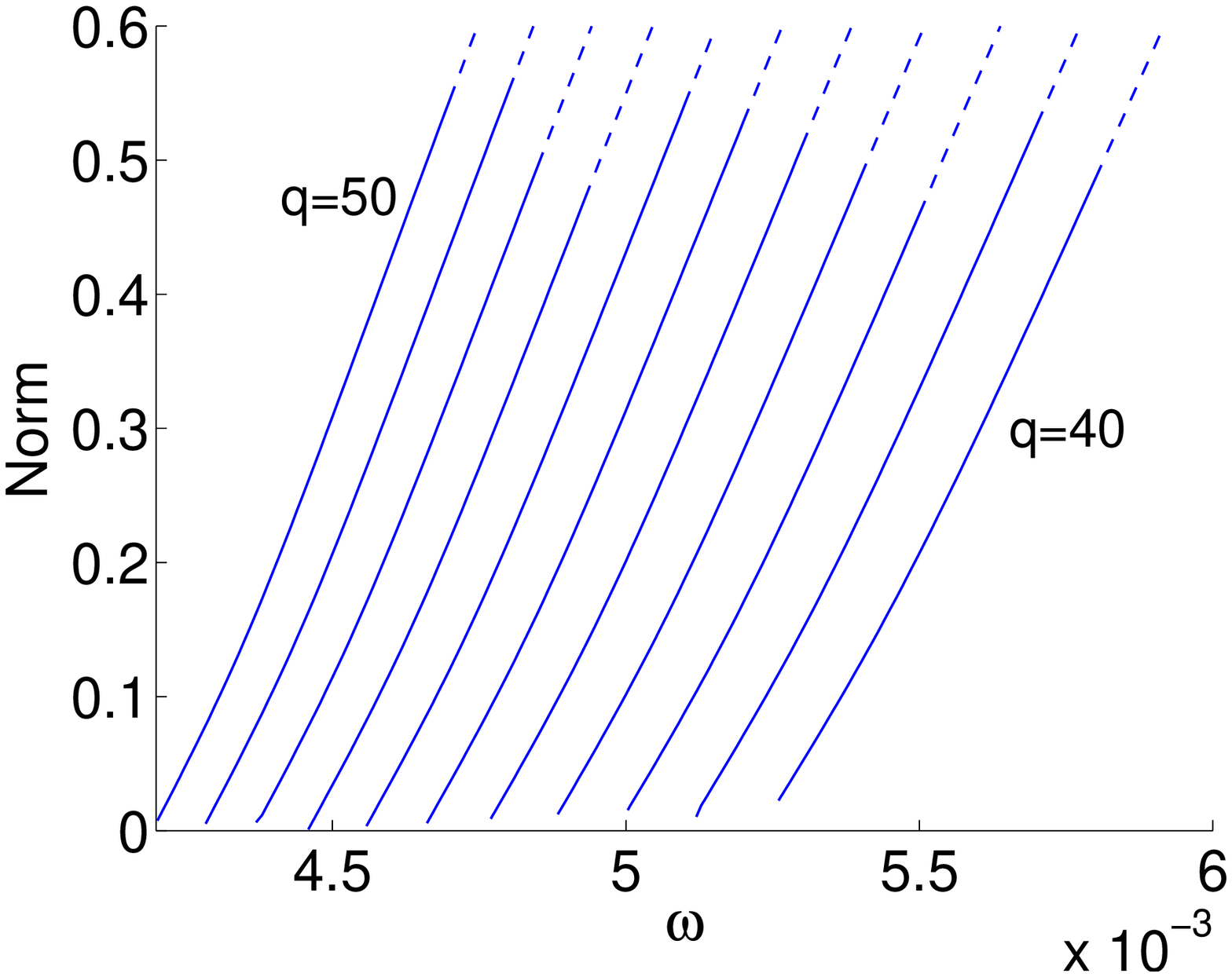}
    \includegraphics[width=7cm]{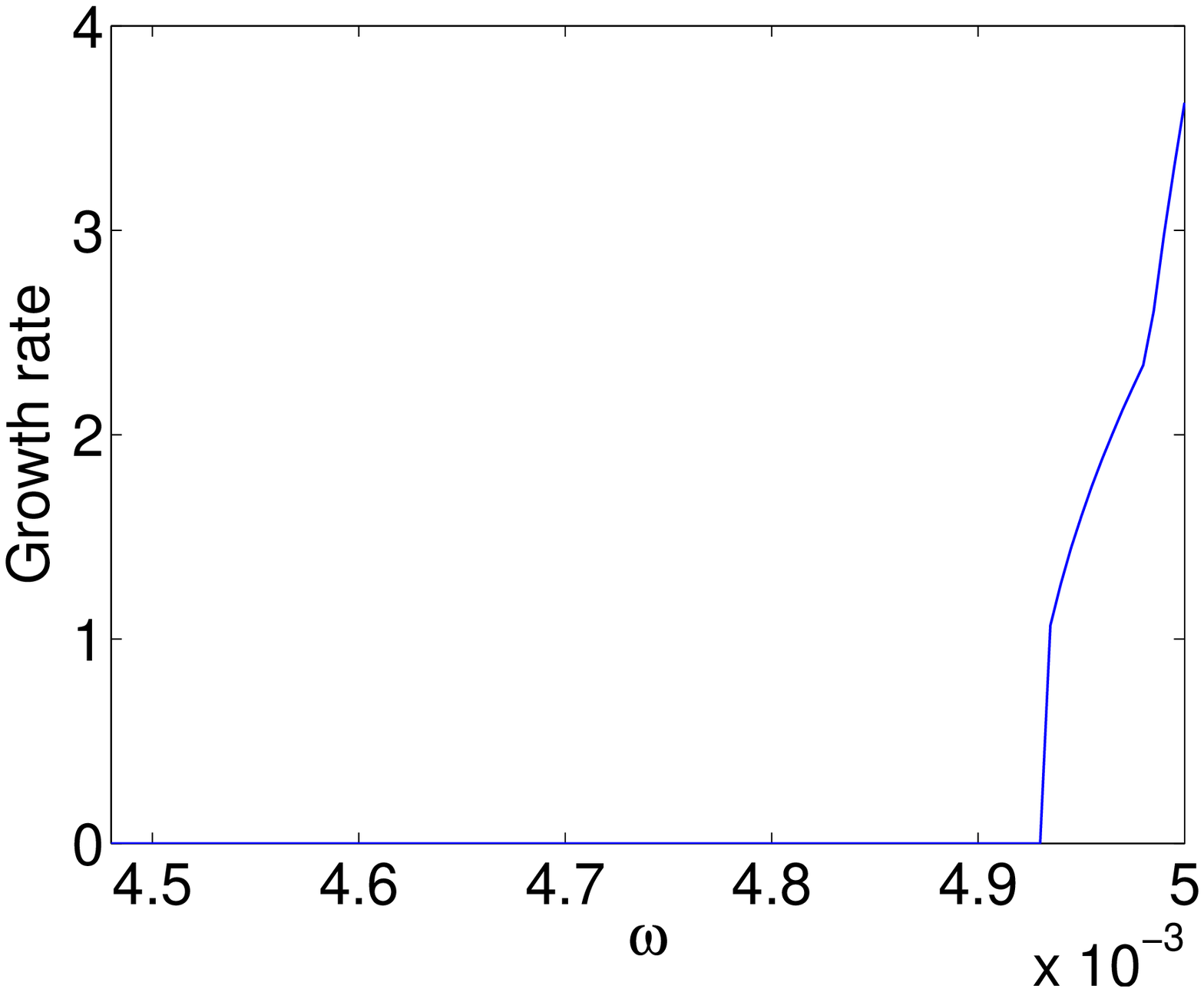}\\
        \includegraphics[width=7cm]{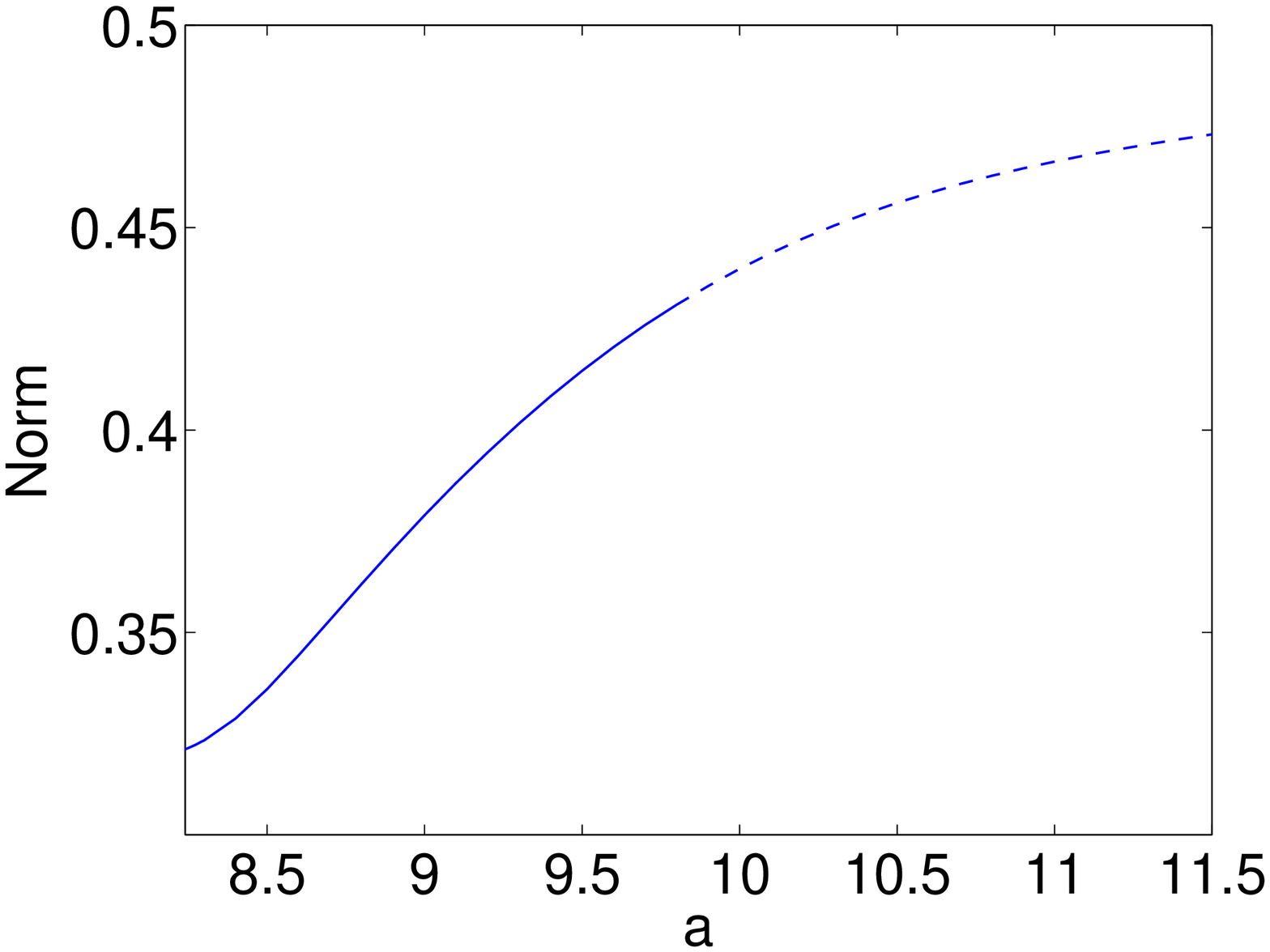}
    \includegraphics[width=7cm]{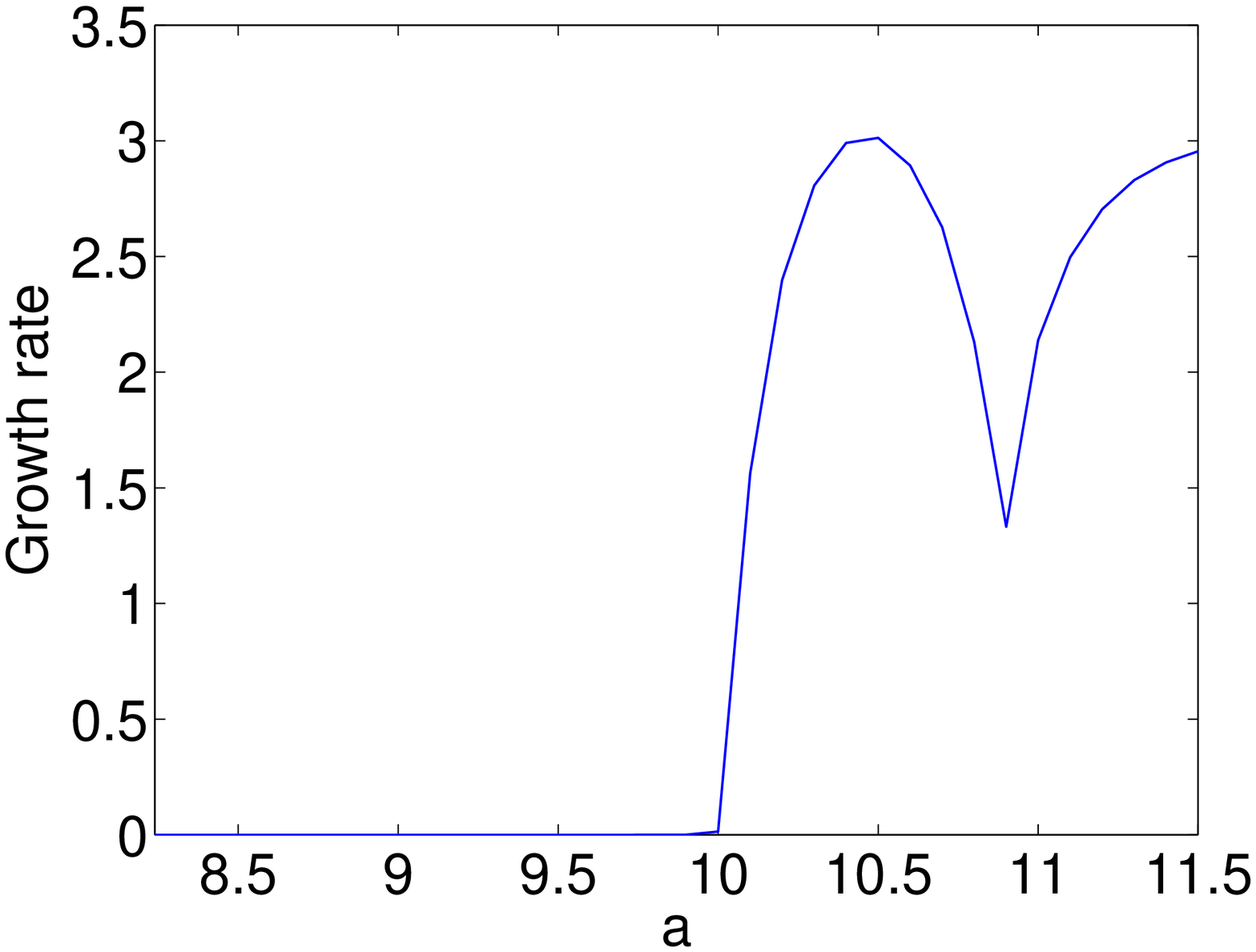}
\caption{The top left panel presents the dependence of the norm with respect to $\omega$ for dark solitons with $a=10$. 
Dashed lines indicate unstable solutions. Here, $q$ sweeps the values between 40 and 50. The top right panel shows the dependence of the growth rate with respect to $\omega$ for $q=47$. Bottom panels depict the norm and the growth mode of tunneling dark solitons with fixed $\omega=0.0049$ and $q=47$ for varying $a$. }
\label{fig:norm}
\end{figure}

We have also obtained periodic solutions for various parameter values. In the top left panel of Figure \ref{fig:norm} we show the dependence of the norm (number of atoms) $N$ of tunneling dark solitons when the inter-well oscillation frequency is varied. In the panel, several representative values of $q$ are considered. Note that the possible values of $q$ are not limited to those shown in the graph. As $\omega$ is increased further, there is a critical value above which solutions are unstable. Unstable solutions are indicated as dashed line in the top left panel. The solutions can also be continued for decreasing frequencies $\omega$ down to a critical value for which the solutions transform into a non-oscillating one (not shown here). In the top right panel of Figure \ref{fig:norm} we show the dependence of the growth rate (the logarithm of the maximum modulus of the Floquet operator eigenvalues) with respect to $\omega$ for $q=47$. We also present the growth rate of JT for a fixed $\omega$ and $q$ and varying separation distance between the two wells $a$ in the bottom panels of the same figure, i.e.\ $\omega=0.0049$ and $q=47$. For small $a$, the solutions tend to a non-oscillating one with one dark soliton in each well, analogously to what occurs for small $\omega$ and fixed $a$.

\section{Coupled-mode approximations and their validity}

\begin{figure}[tbp]
\includegraphics[width=8cm,angle=0,clip]{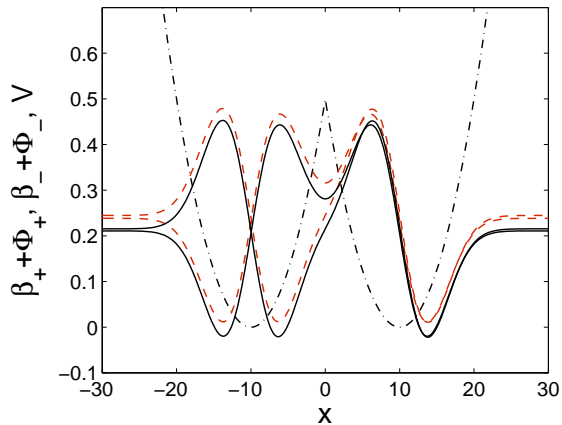}
\caption{The second  and third  collective modes of the confining potential $V(x)$ (dash-dotted) for (solid) $N=0.034$ and (dashed) $N=0.7677$, with $s=-1$. 
}
\label{fig:coeffs}
\end{figure}

To describe dark soliton dynamics reported in the previous section, we will readily use a two-mode approximation derived in \cite{anan06}. Following \cite{anan06}, we write
\begin{equation}
\psi=\sqrt{N}\left(b_{2}(t)\Phi_{2}(x)+b_{3}(t)\Phi_{3}(x)\right), \, \Phi_{2,3}=\frac{\Phi_+(x)\pm\Phi_-(x)}{\sqrt{2}},
\label{ans}
\end{equation}
where $\Phi_{\pm}(x)$ is a continuation of $\phi_{2,3}$ (\ref{linmod}) for nonzero $a$ satisfying
\begin{equation}
\partial_{xx}\Phi_\pm+\beta_\pm\Phi_\pm-V(x)\Phi_\pm+sN\Phi_\pm^3=0,
\end{equation}
with the constraint $\int_{-\infty}^{+\infty}\Phi_j\Phi_k\,dx=\delta_{j,k}$, $i,j=+,-$. Two examples of $\Phi_j$, which corresponds to the norm $N$ in the Figure \ref{fig:evol} are presented in Figure \ref{fig:coeffs}.

Next, for simplicity we write $b_j(t)=|b_j(t)|e^{i\theta_j(t)}$. Equations (\ref{nor}) and (\ref{ans}) imply that $|b_2(t)|^2+|b_3(t)|^2=1$. Defining
\begin{equation}
z(t)=|b_2(t)|^2-|b_3(t)|^2,\qquad \Delta\theta(t)=\theta_3(t)-\theta_2(t),
\end{equation}
one can obtain the equations satisfied by $z$ and $\Delta\theta$ \cite{anan06}
\begin{equation}
\frac{dz}{dt}=-\frac{\partial H}{\partial\Delta\theta},\qquad\frac{d\Delta\theta}{dt}=\frac{\partial H}{\partial z},
\label{pp}
\end{equation}
where
\begin{eqnarray}
H=\frac12Az^2-B\sqrt{1-z^2}\cos\Delta\theta+\frac12C(1-z^2)\cos 2\Delta\theta,\\
A=\frac{10\gamma_{+-}-\gamma_{++}-\gamma_{--}}4,\quad B=\beta_--\beta_++\frac{\gamma_{++}-\gamma_{--}}2,\\
C=\frac{-2\gamma_{+-}+\gamma_{++}+\gamma_{--}}4, \quad \gamma_{jk}=-sN\int_{-\infty}^\infty \Phi_j^2(x)\Phi_k^2(x)\,dx,\\
\end{eqnarray}
with $j,k=+,-$. 

\begin{figure}[tbp]
\includegraphics[width=7cm,angle=0,clip]{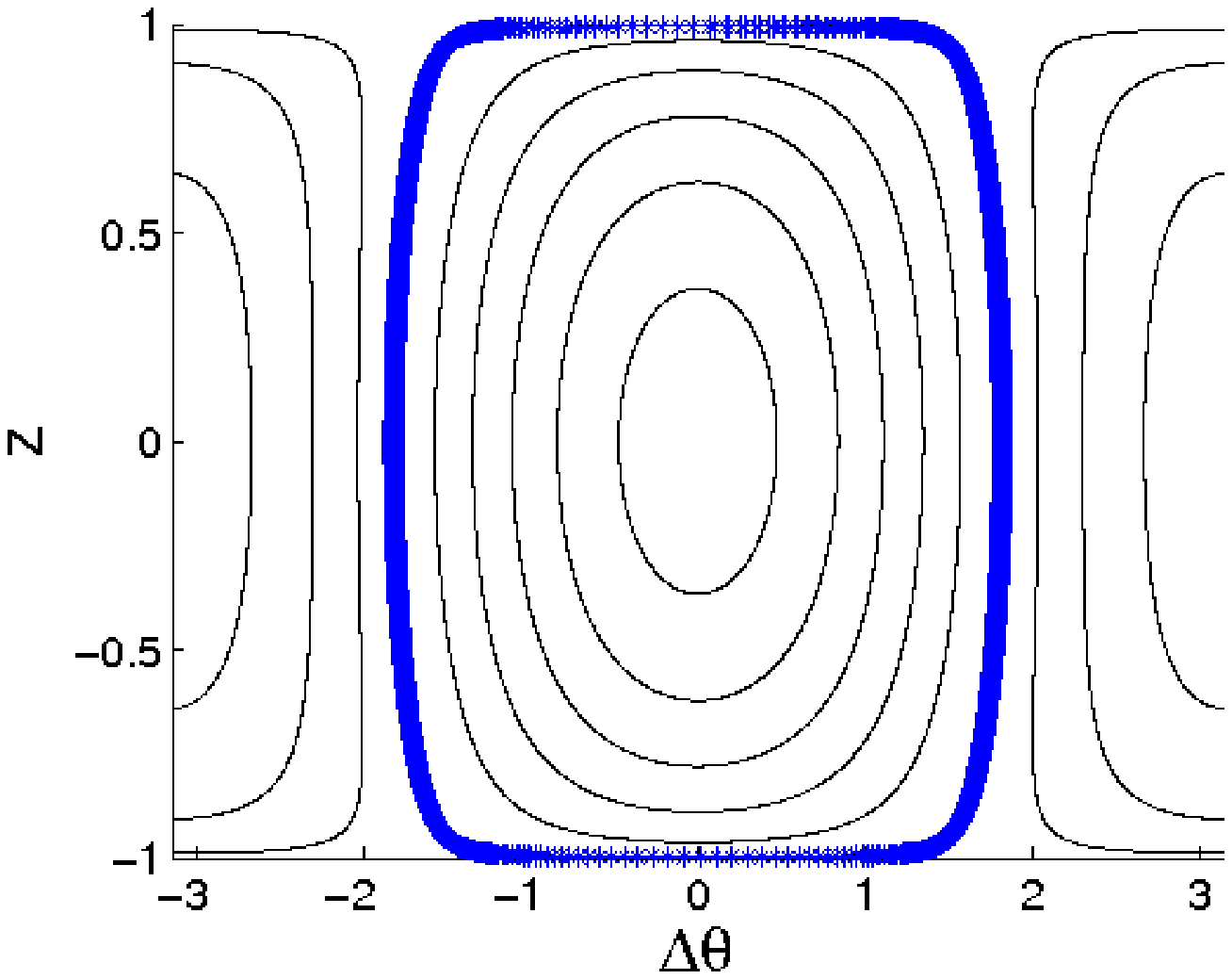}
\includegraphics[width=7cm,angle=0,clip]{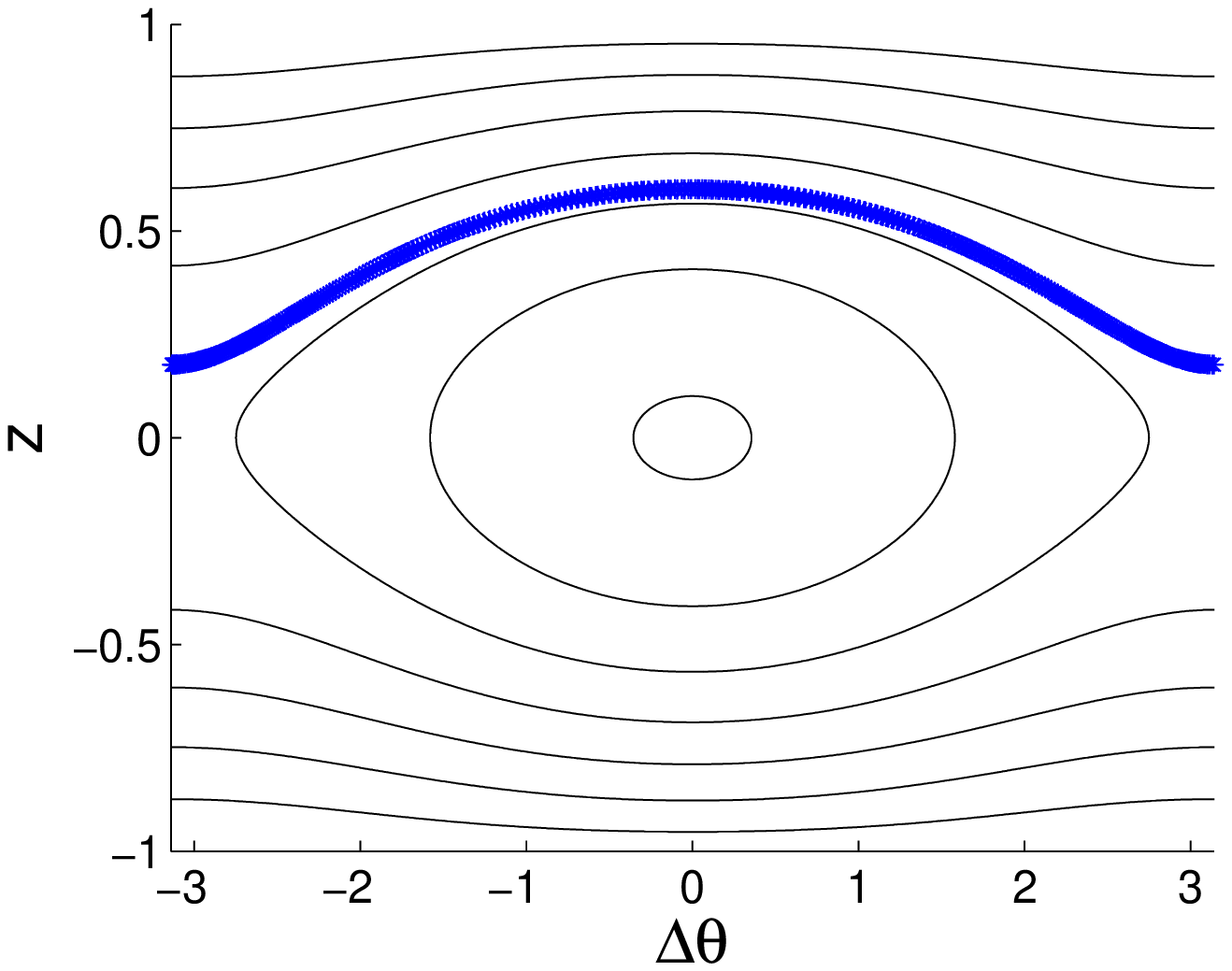}
\caption{The phase-portraits of (\ref{pp}) for the two values of $N$ in Figure \ref{fig:evol}, i.e.\ (left) $N=0.0340$ and (right) $N=0.7677$. Thick symbols correspond to the periodic solutions shown in Figure \ref{fig:evol}.}
\label{pp2}
\end{figure}

We plot the phase-portrait of (\ref{pp}) in Figure \ref{pp2} for the two values of $N$ in Figure \ref{fig:evol}. To compare the two-mode approximation with the top panels of Figure \ref{fig:evol}, we calculate the variable $z$ from the numerics of the full equation (\ref{ge1}) as \cite{anan06}
\[
z=\frac{\int_{-\infty}^0|\psi(x,t)|^2\,dx-N/2}{NS},\quad S=\left|\int_{-\infty}^0\Phi_+\Phi_-\,dx\right|,
\]
where in the present case $S\approx0.5$. 
As $\Delta\theta$ can be calculated immediately, one can compare the numerics and the approximation right away. Shown in Figure \ref{pp2} are the comparisons, where satisfactory agreement is obtained. As for the instability of the solution in the top right panel, that develops at a later time, it is beyond the validity of any currently available two-mode approximations. One needs a better ansatz for the approximations to capture the stability of the periodic solutions. Note that the validity issue discussed herein is completely different from that in \cite{anan06}. In \cite{anan06}, the issue is related to the fact that the approximation does not capture the Josephson oscillation of the full equation directly from the beginning, which typically occurs when $|sN|\gg1$, while in our case $|sN|<1$ and the approximation does capture the existence, but not the stability.

One can observe that the phase portrait in the left panel of Figure \ref{pp2} has two families of periodic oscillations, i.e.\ one centred at $\Delta\theta=0$ and the other at $\pm\pi$. The latter is known as $\pi$-oscillations \cite{ragh99}. The stable solution in the top left panel of Figure \ref{fig:norm} with $q=50$ and the same norm belongs to this family. With this, we conjecture that all trajectories in the phase portrait are stable.

For the phase portrait in the right panel of Figure \ref{pp2}, one can observe that there are two types of solutions, i.e.\ Josephson oscillations and running states. The latter type is also referred to as macroscopically quantum self-trapped states. As we have seen that the two-mode approach can provide a good approximation to periodic solutions, one can use the ansatz (\ref{ans}) and direct simulations of (\ref{ge1}) to predict whether or not a periodic solution is stable. With this, we conjecture that all trajectories in the phase portrait are unstable.

Based on our analysis above, it is likely that all periodic solutions belonging to the same phase portrait, i.e.\ having the same norms, will have the same stability property. Nonetheless, one can easily notice in the top left panel of Figure \ref{fig:norm} that the critical norms, above which the corresponding solutions are unstable, for different values of $\omega$ are not exactly the same. Further analysis whether or not the discrepancy is caused by the finite number of Fourier modes $k_m$ is to be addressed in future investigations.

\section{Experimental setup}

Dark solitons can be created in a magnetic trap in various ways, including phase-imprinting \cite{burg99,beck08,stel08}, merging of two condensates \cite{well08}, and passing a penetrable barrier through condensates \cite{enge07}. 
In the context of condensate splitting and merging, a method of applying oscillating radio-frequency (rf) fields in combination with static magnetic fields has been recently proven to facilitate good experimental control over creating, tuning, and manipulating double well potentials \cite{Schumm,Lesanovsky,Hofferberth}. Without changing the static field configuration that provides to very good approximation a parabolic single well trap, for example in atom chip based microtraps \cite{Folman}, modifications of frequency and amplitude of the rf field allow for raising and lowering a splitting potential barrier. The barrier height between the wells is readily controlled in this type of setup. Moreover, inhomogeneities of the rf field, introduced by locally producing the field by an additional conductor on an atom chip \cite{Schumm}, can be exploited to introduce an imbalance or slight asymmetry between the two wells. Note that rf field engineering in the context of microtraps with typical distances between trapped atoms and field sources on the order of $10-100\,\mu m$ is straightforward as the wavelength of the oscillating field (in the MHz range) is by far sufficiently long to warrant a near-field treatment, with a DC calculation yielding accurate results.

\begin{figure}[tbp]
\centerline{
\includegraphics[width=10cm,angle=0,clip]{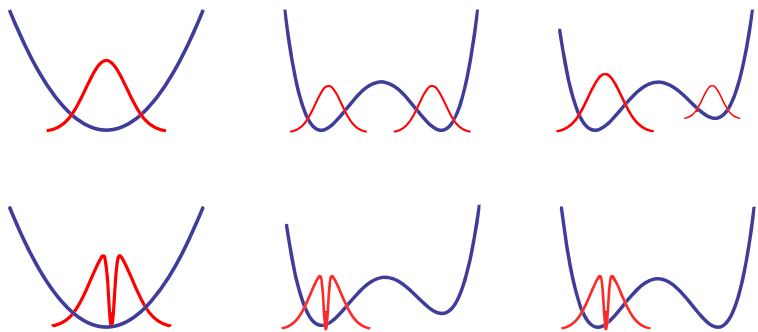}
}
\caption{Schematic implementation of a protocol for controlled production of a soliton in a double well potential (see text).}
\label{fig6}
\end{figure}

Figure \ref{fig6} illustrates a possible scheme based on these methods to produce a soliton in an originally harmonically trapped degenerate Bose gas and to observe its subsequent Josephson oscillations in a double well potential. The protocol starts with a simple harmonic magnetic trap, as can be produced on an atom chip by a Z-shaped wire \cite{Folman}. The gas is then split with rf fields into a symmetric double well. The double well is then slightly imbalanced so that after a time of typically a few ms, a relative phase difference of the two clouds of $\phi=\pi$ will have accumulated. Merging the potentials will now result in a solitonic excitation of the combined gas (see, e.g., \cite{negr04}). Slight controllable deviations $\Delta\phi$ from $\pi$ will produce a (slowly) oscillating soliton. After raising the barrier once again, the tunneling dynamics of the soliton can now be studied. The scheme can be extended to multiple solitons by splitting the potential in more than two wells, for example by applying multiple frequency component rf fields \cite{lesa07}, relevant in the study of interactions of different numbers of dark solitons in each well.

\section{Conclusion and future work}

We have studied dark soliton dynamics in a double well potential, where it has been shown that dark solitons can experience JTs between the wells. A coupled-mode approximation has been derived to explain the observations. 
Numerical stability analysis based on the full governing equation has been performed to show that JTs can be stable. Through path following methods, unstable solutions were also obtained. 
An experimental scheme designed to explore such dark soliton dynamics in the laboratory also has been outlined.

\begin{figure}[tbp]
\centerline{
\includegraphics[width=7cm,angle=0,clip]{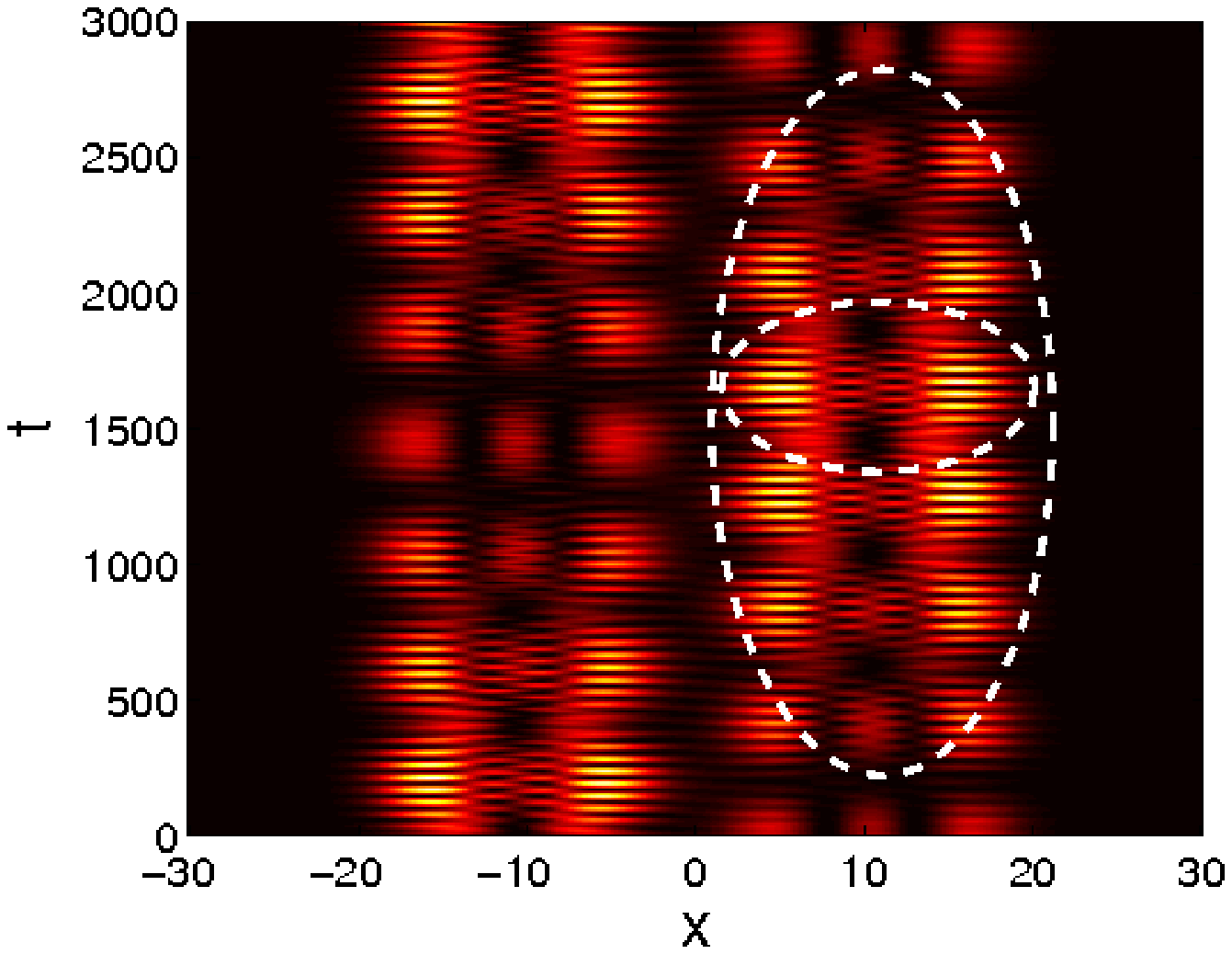}
\includegraphics[width=6.7cm,angle=0,clip]{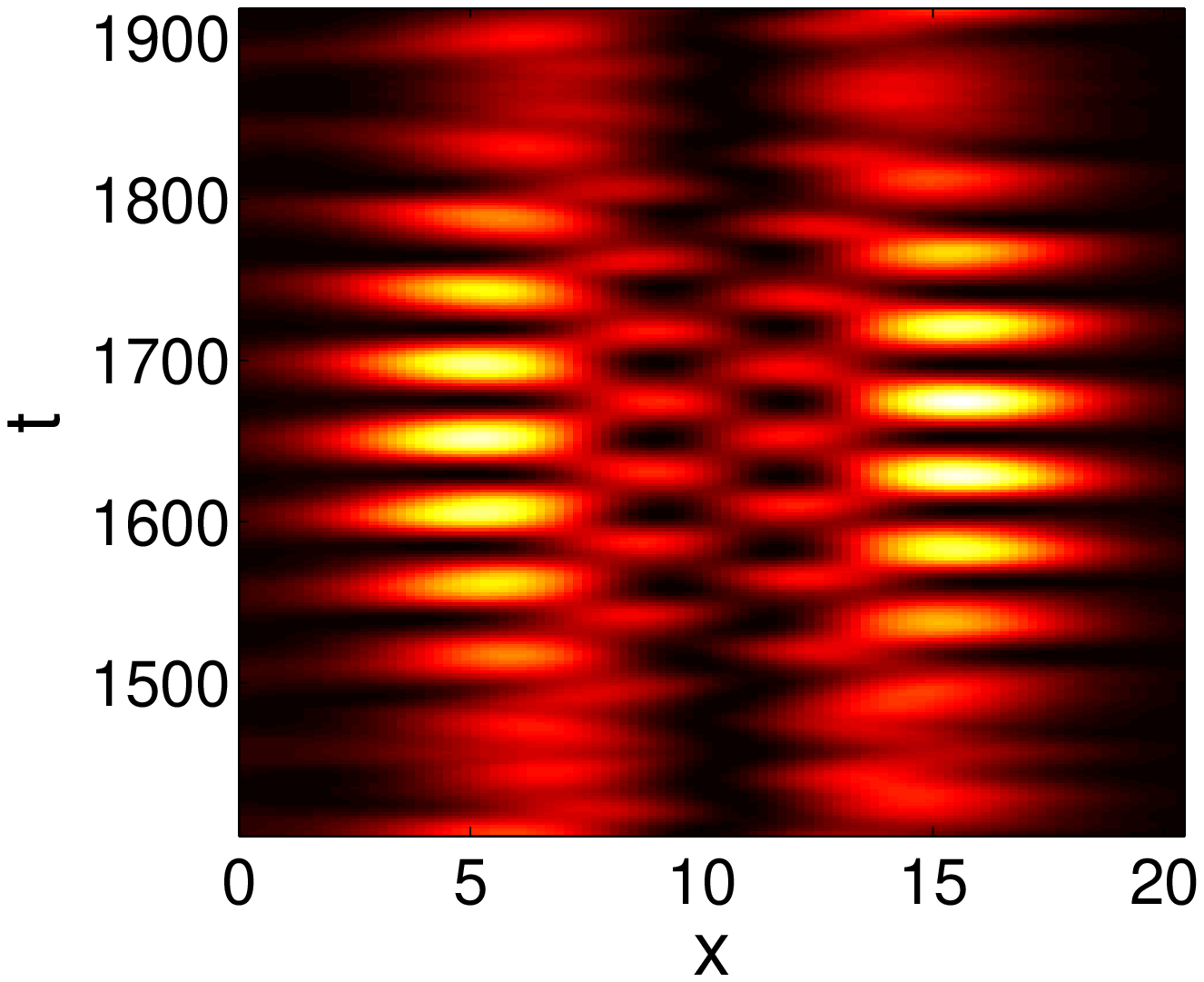}
}
\caption{(Color online) Interactions between two different excitations when loaded into each of the minima of the magnetic trap. Big and small dashed ellipse indicates one tunneling period of the excitation in the left and right well, respectively. The bottom panel zooms in the small dashed ellipse. The total number of atoms is small, similar to that in Figure \ref{fig:evol}. 
}
\label{fig3}
\end{figure}

A natural problem to follow the tunneling reported above is when two different numbers of dark solitons are loaded into each of the minima. We present in Figure \ref{fig3} interactions between one and two dark solitons. One can observe that there is an interference pattern 
analogous to the acoustic beating pattern in the interaction of two continuous waves with slightly different frequencies. As shown in the left panel indicated by the dashed ellipses, there are two levels of modulated patterns; the big ellipse shows one tunneling period modulated by the oscillation in the small ellipse. In the right panel, we zoom in on the small ellipse to show that a beating pattern also occurs on a smaller scale. It can be calculated that the oscillation period in the small ellipse when $a\gg1$ is approximately $T\approx2\pi/\Omega$, which is in accordance with the numerical result. A multi-mode approximation can be obtained as before as briefly discussed in \cite{anan06}. It is then interesting to study the stability of such  interactions. Together with the question on a better ansatz that can predict the stability of periodic solutions reported herein, this is currently under investigation and will be reported elsewhere.

\ack
We acknowledge fruitful discussions with Andrea Trombettoni and Panos Kevrekidis. JC acknowledges financial support from the MICINN project FIS2008-04848.

\end{document}